\documentclass[10pt]{article}
\pdfoutput=1
\usepackage{graphicx, verbatim, url, amssymb, amsmath, amsfonts, amsthm, latexsym, enumerate, float, subcaption}

\usepackage{booktabs}
\usepackage{cool}
\usepackage{graphicx}
\usepackage{multirow}
\usepackage[square]{natbib}
\usepackage[utf8]{inputenc}
\usepackage[english]{babel}

\begin{document}
	
\title{A diachronic study of historiography\footnote{This research has been funded in part by the Swiss National Fund with grants 205121\_159961 and P1ELP2\_168489.}
}

\author{Giovanni Colavizza\\
	The Alan Turing Institute\\
	\url{gcolavizza@turing.ac.uk} 
}

\date{}

\maketitle

\begin{abstract}
	The humanities are often characterized by sociologists as having a low mutual dependence among scholars and high task uncertainty. According to Fuchs' theory of scientific change, this leads over time to intellectual and social fragmentation, as new scholarship accumulates in the absence of shared unifying theories. We consider here a set of specialisms in the discipline of history and measure the connectivity properties of their bibliographic coupling networks over time, in order to assess whether fragmentation is indeed occurring. We construct networks using both reference overlap and textual similarity. It is shown that the connectivity of reference overlap networks is gradually and steadily declining over time, whilst that of textual similarity networks is stable. Author bibliographic coupling networks also show signs of a decline in connectivity, in the absence of an increasing propensity for collaborations. We speculate that, despite the gradual weakening of ties among historians as mapped by references, new scholarship might be continually integrated through shared vocabularies and narratives. This would support our belief that citations are but one kind of bibliometric data to consider --- perhaps even of secondary importance --- when studying the humanities, while text should play a more prominent role.
\end{abstract}

\section{Introduction}\label{sec:intro}

Diachronic change is a crucial aspect of the intellectual organization of any scholarly community. Philosophers and sociologists of science have amply reasoned about this phenomenon. Perhaps the most well-known theory of change in science is Kuhn's paradigm shift \citep{kuhn_structure_1996}, where science is pictured as developing through a sequence of perception-altering revolutions instating new paradigms of scientific inquiry destined to last for longer periods of normal science. Some sociologists, including \cite{collins_conflict_1975}, \cite{whitley_intellectual_1984} and \cite{fuchs_professional_1992}, have instead emphasized two other aspects influencing the way scientific communities change, among others: the degree of mutual dependence among scholars and the degree of task uncertainty \citep[Ch. 2]{chen_representing_2017}. These influence, in particular, the possible development of a community towards increased intellectual \textit{specialization} or \textit{fragmentation}, as new results are produced. According to Fuchs' theory of scientific change \citep{fuchs_sociological_1993}, we predict a community to develop towards increased specialization if it possesses a high degree of mutual dependence among scholars, for example via a guiding theoretical framework, a shared set of questions or a centralized research infrastructure, and a low degree of task uncertainty, so that most tasks are relatively repetitive and their outcomes predictable. Conversely, a fragmenting community possesses a low degree of mutual dependence and a high degree of task uncertainty, such is the case, according to Fuchs, for most of the social sciences and the humanities.

The problem of the effects of the accumulation of new literature in historiography is often discussed by historians as one of (perceived) \textit{over-specialization}, thus with a negative connotation. In particular, during and after periods of sustained growth, scholars tend to lament a raise of technical specialization as manifested by the narrowing focus of new publications, and the effects this has on the intellectual fragmentation of the field and on the scope of questions the community considers \citep{tyrrell_historians_2005,colavizza_hoh_2018}. More to our point, in bibliometrics it remains an open question how scholars, in particular in the humanities, react and adapt to the vertiginous amount of literature currently being produced \citep{bornmann_growth_2015}, and to its raising digital discoverability and availability \citep{evans_electronic_2008,lariviere_decline_2009}. In this article we therefore focus on the case study of history as a discipline, and propose a measure of the degree of cohesiveness of its intellectual organization. We then consider if this degree is changing over time. 

There are some preliminary design choices to make when studying a research field in this perspective, first of all the scale of analysis. By considering a single, sufficiently small community, it might be possible to analyze a longer span of time at a more granular level. Another option is to perform a large-scale analysis relying on databases such as the Web of Science (WoS) or Scopus, which would guarantee to consider (many) more observations, but over a shorter span of time and with an exclusive focus on journal articles. We take an intermediate path here, by considering five specialisms: economic history, social history, history of science, history of medicine and general English history. We do so by using data from WoS, representing each specialism using a set of three journals each, considered from the early 1950s to 2016 included, as allowed by data availability. Despite the focus on journal articles as citing publications, we consider citations to both source and non-source items (i.e. indexed in WoS or not). The focal point of attention will be the bibliographic coupling networks of each specialism so represented, considered over subsequent intervals of time. In particular, we focus on \textit{network connectivity} as a proxy for the degree of cohesiveness of these specialisms, in an attempt to detect whether connectivity is raising, stable or declining over time. Three bibliographic coupling networks will be considered: i) of article citations (or reference overlap), ii) of author citations (identical but considering authors as nodes instead of articles), iii) of article textual similarity (measured using each article's title and abstract). We thus use bibliographic coupling in a generic way, to name networks where the nodes are scholarly publications or authors, and the edges are determined by some similarity measure among any two nodes.

This article starts by a brief overview of the state of the art and discussion of Fuchs' theory of scientific change. We will then introduce the methods used to construct the afore-mentioned bibliographic coupling networks, and those used to compare these networks over time, measuring network connectivity. Next, data and results are presented and discussed, while a summary with a future outlook is given in the conclusions.

\section{State of the art}\label{sec:soa}

The humanities are considered to possess a set of characteristics which make it more challenging to acquire and use citation data to study their intellectual organization and communication practices. Nevertheless, the available evidence leads us to suspect that the accumulation of knowledge in the humanities might follow patterns distinct from the sciences (for what follows see, among others: \cite{garfield_is_1980,glanzel_bibliometric_1999,barrett_2005,van_leeuwen_application_2006,hellqvist_referencing_2009,linmans_why_2009}. For reviews see instead: \cite{hicks_difficulty_1999,nederhof_2006,huang_characteristics_2008}). First of all, humanists often strongly feel the importance of the national and local dimensions in terms of their reading public, the language they use to publish and the object of their research. For a historian, for example, it is impossible to abstract from place and time, which often entail the need to access primary sources which are specifically located and written in local languages. Humanists use a variety of publication typologies and are not just focused on journal articles. It is worth stressing that monographs are especially important, as the practice in the humanities still favors them over other kinds of publications in order to get recognition within the field, despite variations in citation patterns among different disciplines \citep{knievel_2005,williams_role_2009}. The humanities have also been found to identify core works at a slower pace than other sciences, in part due to longer times required for citation accumulation, entailing a long life-span of publications \citep{nederhof_2006}. The referencing practices of humanists are also syntactically and semantically richer than in the sciences. For example, recent work has found a significant amount of perfunctory (non-essential) and negative citations in the literature of historians \citep{lin_citation_2013}. The broad variety of topics and sources being investigated also results in a less focused and wider information retrieval behavior. Manual reference chaining remains important \citep{buchanan_2005}, and browsing is particularly needed for historians, especially so in archives \citep{talja_reasons_2003}. The research library remains thus central for locating literature \citep{stone_humanities_1982}. More recently keyword search, used online and on catalogs or tools such as Google Books, has become more popular \citep{fry_intellectual_2007}. Humanists indeed increasingly depend now on a myriad of digital tools, which they often use as non-advanced users \citep{trace_information_2017}. Lastly, at the level of their social behavior, humanists have in general little propensity for collaboration and team-work, as attested by co-authorship patterns \citep{kyvik_research_2017}. These considerations in part motivate why, so far, ``the study of the intellectual structure within the humanities using citation analysis is as yet an underdeveloped area'' of research \citep{hammarfelt_2011}. Recent work highlighted some commonalities to be found in the literature on the topic \citep{colavizza_structural_2017}:
\begin{enumerate}
	\item The reliance on existing citation indexes, above all the A\&HI.
	\item The almost lack of general maps, but instead a focus on disciplines or fields of research.
	\item The presence of several attempts to counter the lack of data, e.g. by using non-source items.
	\item The still immature state of theoretical developments on the intellectual organization of the humanities.
\end{enumerate}

The present work, while sharing traits 1 to 3, attempts to tie empirical results with theoretical developments from related fields. Several sociologists of science, in particular, characterize the humanities separately from the sciences, in part due to the reasons discussed above. A clear sociological and organizational underpinning informs \cite{fuchs_sociological_1993}'s theory of scientific change, part of his broader theory of scientific organizations (TSO) \citep{fuchs_professional_1992}. The TSO views research specializations as ``reputational work organizations'', as in \cite{whitley_intellectual_1984}. There are three aspects to the TSO: a sociological perspective stating that the ways we think and perceive the world are shaped by social structures (e.g. the degree of mutual dependence among scholars); that social and cognitive structures are also informed by the way work is done and technology is used (e.g. the degree of task uncertainty and the role of research technologies); and a materialist theory of consciousness by which the ways we think are related to the control of the ``material means of mental production'' (e.g. the centralization of technologies in '`big science' projects). A central tenet of the theory are the (recurring) axes of mutual dependence and task uncertainty, borrowed from \cite{collins_conflict_1975} and \cite{whitley_intellectual_1984}. Fuchs' most significant contribution is a focus on scientific change and its explanation through \textit{competition}.

\begin{figure}
	\centering\includegraphics[width=8cm]{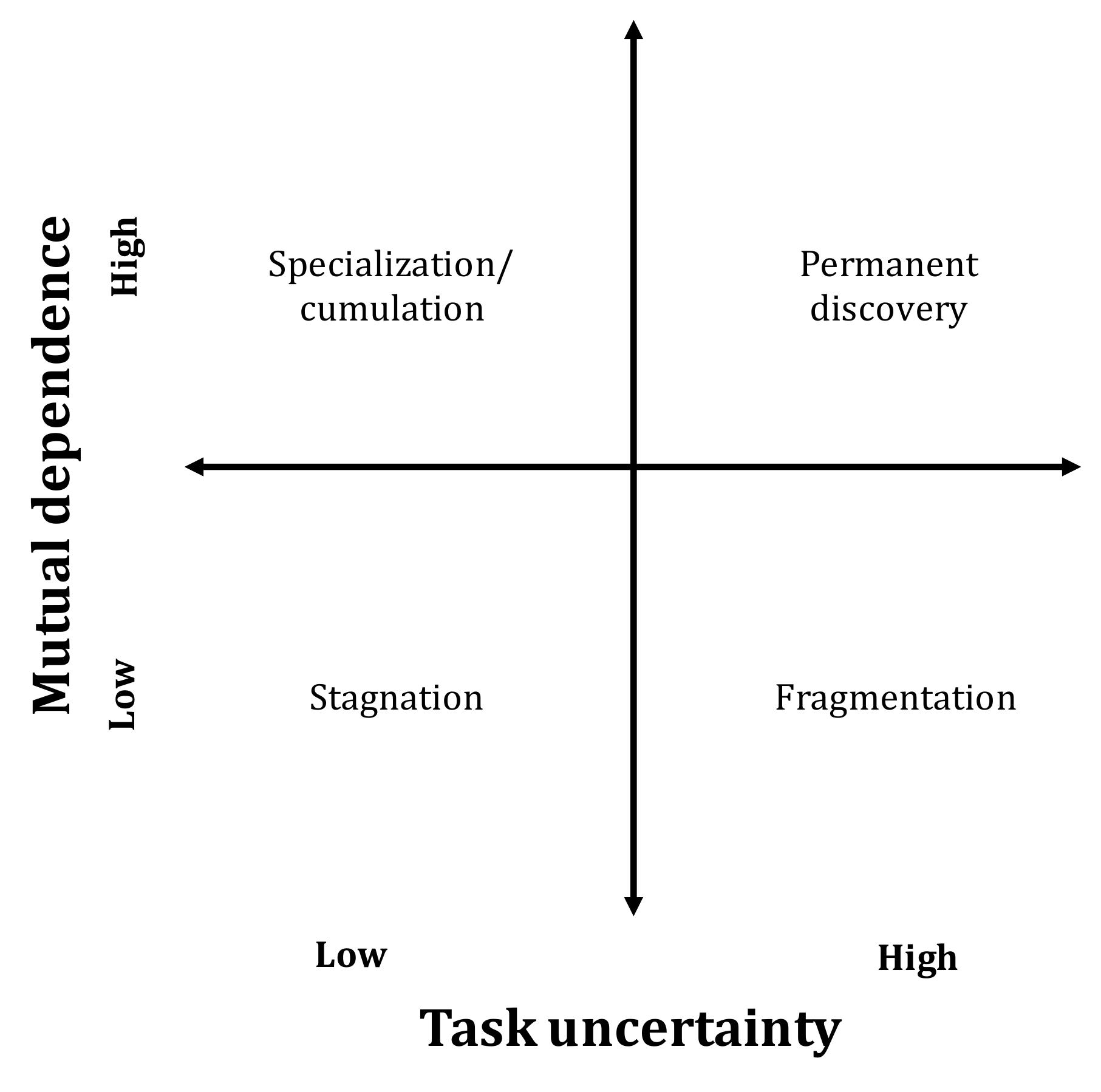}
	\caption{Three types of scientific change \citep[940, Figure 1]{fuchs_sociological_1993}.}
	\label{fig:fuchs}
\end{figure}

Fuchs argues that, within ``reputational work organizations'' such as academic research fields, competition is a major drive of change, and the way competition acts varies according to the two axes of mutual dependence and task uncertainty, as shown in Figure \ref{fig:fuchs}. The three main effects of competition are: to foster a state of permanent discovery in fields which possess both high mutual dependence and task uncertainty; to instead push towards increasing specialization and knowledge cumulation in fields with high mutual dependence but low and predictable task uncertainty; finally, to generate fragmentation in fields with low mutual dependence but high task uncertainty, such is often the case in the humanities. Eventually, fields in a regime of both low task uncertainty and mutual dependence are seen as stagnant. An important insight is that different ways of doing science actually coexist and are practiced by different cohorts of researchers within fields; most notably a group of leaders can proceed by permanent discovery at the core while many more researchers work cumulatively at the periphery of a community.

Most relevant for our purposes is the distinction between specialization and fragmentation. Essentially, both are effects of similar strategies to cope with competition and, it will be argued, information load. The different outcomes relate to the presence or absence of a ``paradigmatic integrity in the larger field'' which can be used to effectively integrate new contributions in the larger body of knowledge mostly via a shared theoretical framework, possibly embedded in research technologies and methods. Fragmentation ensues from similar premises but in the absence of such unifying framework.

\section{Methods}\label{sec:methods}

We start by defining the construction of the bibliographic coupling networks. Take $B = (V,E,w)$, the weighted bibliographic coupling network made of the publications of a given specialism, where $w:E \rightarrow \mathbb{R}^+$ is a function mapping each edge to a positive weight. $W$ is the weighted symmetric adjacency matrix representing the graph, where $W_{i,j} = W_{j,i} = w(e_{i,j})$ if there exist an edge between vertices $i$ and $j$, that is to say $e_{i,j} \in E$, $0$ otherwise. Under this general setting, the edges and their weights can be established in a variety of ways. We consider two of them here: reference overlap of articles and authors (i.e. traditional bibliographic coupling as in \cite{kessler_bibliographic_1963}) and textual similarity.

For reference overlap, we consider as the edge weight function $w$ the cosine similarity calculated over the references that two publications $i$ and $j$ have in common:

\begin{align}
w(e_{i,j}) = \frac{R_{i,j}}{\sqrt{R_i}\sqrt{R_j}} \label{eq_cosine}
\end{align}

Where $R_{i,j}$ is the number of references in common between $i$ and $j$, $R_i$ the number of references of $i$. We stress that we consider unique references, not their frequency (number of in-text references, or mentions). The cosine similarity is appropriate as it allows to evenly compare the weight of edges among publications with varied reference list lengths. Author to author bibliographic coupling networks are constructed by considering all (unique) references to publications made by an author within the given specialism and time period.

We base the textual similarity among two papers on the BM25 measure, widely adopted to rank documents for the purpose of information retrieval and document clustering \citep{sparck_jones_probabilistic_2000_1,sparck_jones_probabilistic_2000_2}. This measure has already been applied to assess the textual similarity of scientific publications (e.g. \cite{boyack_clustering_2011,colavizza_closer_2017}), and it improves on simpler tf-idf by explicitly accounting for document lengths. Each publication text -- in our case the concatenation of title and abstract -- is reduced to lower case and split into tokens, further eliminating punctuation and then tokens of just one alphanumeric character. Given a publication $i$ and another publication $j$, the BM25 similarity is calculated as:

$$
s(i,j) = \sum_{z=1}^{n} IDF_z \frac{n_z(k_1+1)}{n_z+k_1\Big(1-b+b\frac{|D|}{|\overline{D}|}\Big)}
$$

where $n$ denotes the number of unique tokens in $i$, $n_z$ equals the frequency of token $z$ in publication $j$, and $n_z = 0$ for tokens that are in $i$ but not in $j$. $k_1$ and $b$ have been set to the commonly used values of 2 and 0.75 respectively. $|D|$ denotes the length of publication $j$, in number of tokens. $|\overline{D}|$ denotes the average length of all publications in the dataset. The $IDF$ value for every unique token $z$ in the dataset is calculated as:

$$
IDF_z = log\Bigg(\frac{N-p_z+0.5}{p_z+0.5}\Bigg)
$$

where $N$ denotes the total number of publications in the dataset and $p_z$ denotes the number of publications containing token $z$. $IDF$ scores strictly below zero are discarded to filter out very commonly occurring tokens. BM25 is not a symmetric measure. We thus obtain a symmetric measure for the similarity of publications $i$ and $j$, the value is the weight of the edge connecting them in $B$, as follows:

\begin{align}
w(e_{i,j}) = \frac{s(i,j)+s(j,i)}{2} \label{bm_25}
\end{align}

While the BM25 textual similarity is calculated for every publication pair, the $IDF$ scores and $|\overline{D}|$, the average length of all publications, are calculated and shared globally over all datasets. We refrain from further normalizing the similarity scores, in order to allow for comparisons across specialisms.

\subsection{Connectivity and giant component}
A connected component of $B$ is a sub-graph whose nodes are all connected, i.e. there exists a path between every pair of nodes in the component. An isolated node is a node that is not connected to any other node (hence representing a singleton connected component). The giant component is the largest connected component in the number of nodes it contains \citep[142-3]{newman_2010}.

In order to explore the connectivity of the bibliographic coupling networks introduced above, we measure the proportion of connected components over the total possible (Eq. \ref{eq1}), at steps in which we remove all edges below a certain weight threshold. This method allows to assess the strength of edge weights in the network, and the behavior of the connected components as the network becomes increasingly disconnected. This procedure can be considered as an analysis of a form of $t$-edge-connectivity, where a component is considered as connected only if it is a connected component by considering edges of weight at least equal to $t$. Alternatively, it is a form of bond percolation where edges are removed deterministically according to their weight. Given an edge weight threshold $t$, we are thus interested in a measure which is calculated at increasing $t$ over networks $B$:

\begin{align}
c(t) = \frac{C^t}{N} \label{eq1}
\end{align}

Where $N$ denotes the number of publications, equivalent to the number of nodes in the network and also equal to the number of connected components in the disconnected network; $C^t$ denotes the number of connected components after removal of edges with weight below $t$. It is worth pointing out that the measure considers the structure of the network after the removal of some edges, but does not account for the weight of the remaining edges. This might appear as a limitation, but in practice the analysis of the process allows to account for both structure and relative weight of edges overall. In fact, alternative measures such as algebraic connectivity or $k$-connectivity \citep{newman_2010}, did not yield complementary results of note.

\section{Data}\label{sec:data}

Five specialisms in history are considered, covering mainstream areas of research. For each, three journals are chosen to represent trends in the research being published therein. Due to the limitations of WoS and in order to maximize coverage over time, the journals were selected to be internationally renown ones in English. An overview of the dataset is given in Table \ref{tab:dataset}. Limitations in the coverage of WoS (which starts at best in 1956), or the date of initial print of some journals, determines an uneven data availability for the initial years under consideration. In any case by the 1980s all journals are active and indexed. With respect of the presence of abstracts, the situation is less fortunate: most journals either do not have them, have them unevenly or they are not present in the database. We therefore consider only one journal per specialism with respect to textual similarity, the one with better abstract availability (marked by an asterisk in Table \ref{tab:dataset}).

The coverage of references allows to consider six time periods for both article and author bibliographic coupling networks: up until 1969 included, 1970-1979, 1980-1989, 1990-1999, 2000-2009, 2010-2016. With respect to textual similarity, only three periods are considered instead: 1999-2004, 2005-2010, 2011-2016. For these reasons, the results relying on textual similarity are to be considered as less robust.

The data were downloaded directly from the WoS interface and processed in order disambiguate all references and authors. Reference lists were first parsed to extract and clean every reference as follows. First, every reference was split into author, publication year and the rest of its text (mainly, the title). The title was then trimmed from any page, number, issue or volume information. Lastly, all anonymous references (in the author field) or references without a publication year were discarded. The second step entailed the creation of a single global reference dictionary for the whole dataset. In order to do this, every reference was compared with every other reference, and a match was established if all the following conditions were verified: a) the first three characters of the author and title fields matched exactly (to lower case); b) the Jaro-Winkler similarity between author fields was equal or greater than 0.9; c) the similarity of the title fields was equal or greater than 0.85 and the publication year matched exactly, or the similarity of the title fields was equal or greater than 0.95 (useful in the case of different editions of the same work). A total of 777'894 references were considered, resulting in 443'561 disambiguated references.

Article authors were instead disambiguated field by field. A match was established if the (lower cased and punctuation stripped) surnames had a similarity strictly higher than 0.95 and names strictly higher than 0.9, using the Jaro-Winkler measure. The number of unique authors per field is given in Table \ref{tab:dataset}, highlighting how the communities of historians of science and of medicine are smaller than the rest, according to our dataset.

\begin{table}[]
	\centering
	\caption{Summary of the dataset. Journals marked with an asterisk are used for creating the text similarity networks, due to their better availability of abstracts.}
	\label{tab:dataset}
	\resizebox{\columnwidth}{!}{%
		\begin{tabular}{|l|c|l|c|c|c|c|}
			\hline
			\textbf{Field}                           & \multicolumn{1}{l|}{\textbf{\# authors}} & \textbf{Journal}                                       & \multicolumn{1}{l|}{\textbf{Coverage}} & \multicolumn{1}{l|}{\textbf{\# articles}} & \multicolumn{1}{l|}{\textbf{Abstracts since}} & \multicolumn{1}{l|}{\textbf{\# articles with abstract}} \\ \hline
			\multirow{3}{*}{General English History} & \multirow{3}{*}{2935}                    & English Historical Review                              & 1956-2016                              & 1220                                      & -                                                      & 3                                             \\ \cline{3-7} 
			&                                          & The Historical Journal*                                & 1966-2016                              & 1784                                      & 1999                                                   & 606                                           \\ \cline{3-7} 
			&                                          & Journal of British Studies                             & 1975-2016                              & 701                                       & 2013                                                   & 117                                           \\ \hline
			\multirow{3}{*}{Social History}          & \multirow{3}{*}{2908}                    & Past\&Present                                          & 1956-2016                              & 1411                                      & -                                                      & 2                                             \\ \cline{3-7} 
			&                                          & Journal of Social History*                             & 1967-2016                              & 1344                                      & 1991                                                   & 737                                           \\ \cline{3-7} 
			&                                          & Social History                                         & 1972-2016                              & 481                                       & 1999                                                   & 138                                           \\ \hline
			\multirow{3}{*}{Economic History}        & \multirow{3}{*}{3399}                    & Explorations in Economic History                       & 1969-2016                              & 1065                                      & 1994                                                   & 544                                           \\ \cline{3-7} 
			&                                          & Journal of Economic History*                           & 1956-2016                              & 1996                                      & 1991                                                   & 734                                           \\ \cline{3-7} 
			&                                          & Economic History Review                                & 1956-2016                              & 1646                                      & 1992                                                   & 639                                           \\ \hline
			\multirow{3}{*}{History of Science}      & \multirow{3}{*}{1872}                    & Isis*                                                  & 1956-2016                              & 1277                                      & 1997                                                   & 406                                           \\ \cline{3-7} 
			&                                          & History of Science                                     & 1987-2016                              & 377                                       & 2014                                                   & 58                                            \\ \cline{3-7} 
			&                                          & Annals of Science                                      & 1966-2016                              & 825                                       & 1991                                                   & 337                                           \\ \hline
			\multirow{3}{*}{History of Medicine}     & \multirow{3}{*}{1727}                    & Bulletin of the History of Medicine*                   & 1977-2016                              & 738                                       & 2001                                                   & 262                                           \\ \cline{3-7} 
			&                                          & Journal of the History of Medicine and Allied Sciences & 1978-2016                              & 550                                       & 2003                                                   & 181                                           \\ \cline{3-7} 
			&                                          & Medical History                                        & 1978-2016                              & 787                                       & 2011                                                   & 136                                           \\ \hline
		\end{tabular}
	}
\end{table}

\begin{figure}
	\begin{minipage}{0.5\textwidth}
		\centering\includegraphics[width=\textwidth]{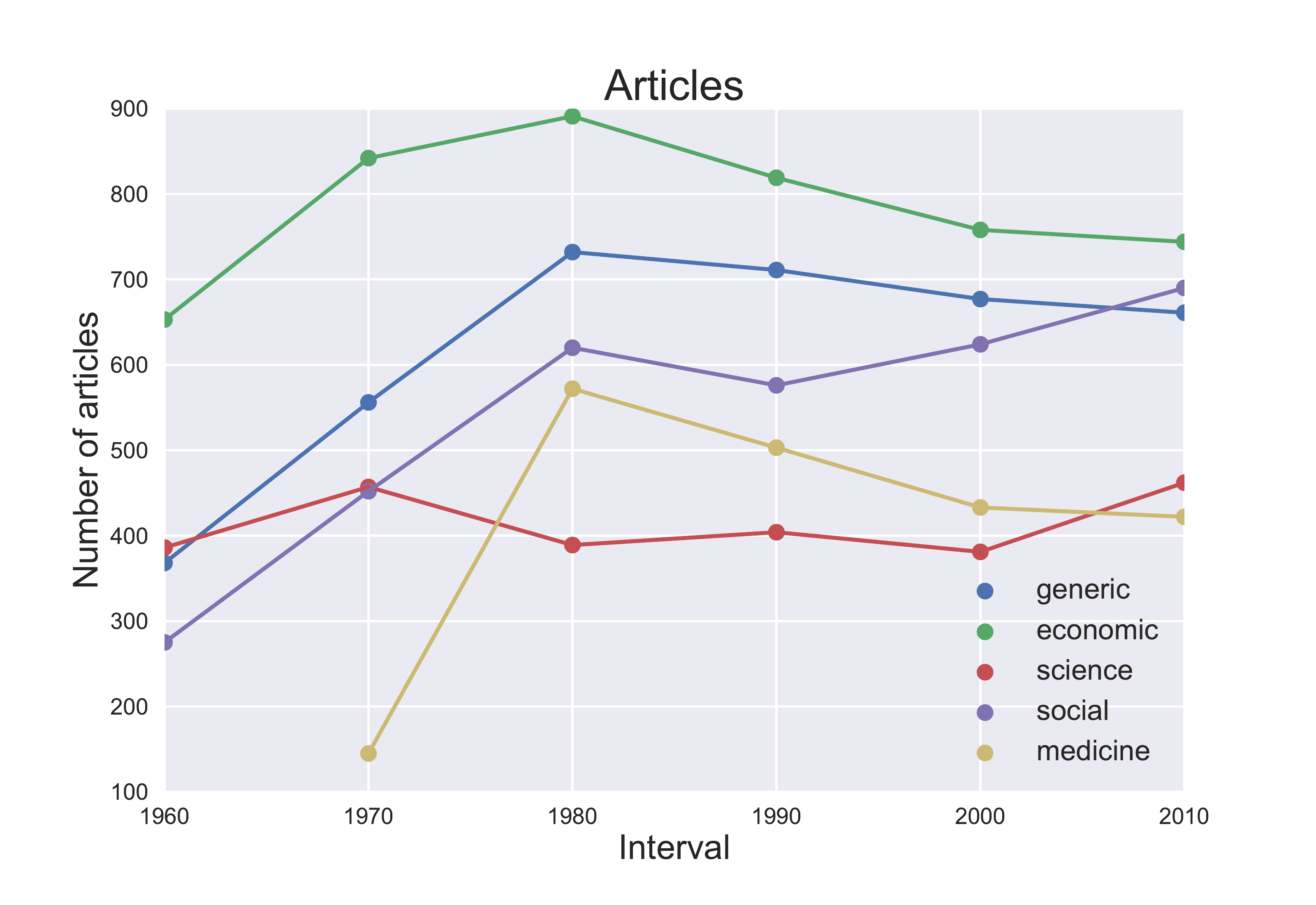}
		\subcaption{Number of articles.}\label{fig:7_n_art}
	\end{minipage}\hfill
	\begin{minipage}{0.5\textwidth}
		\centering\includegraphics[width=\textwidth]{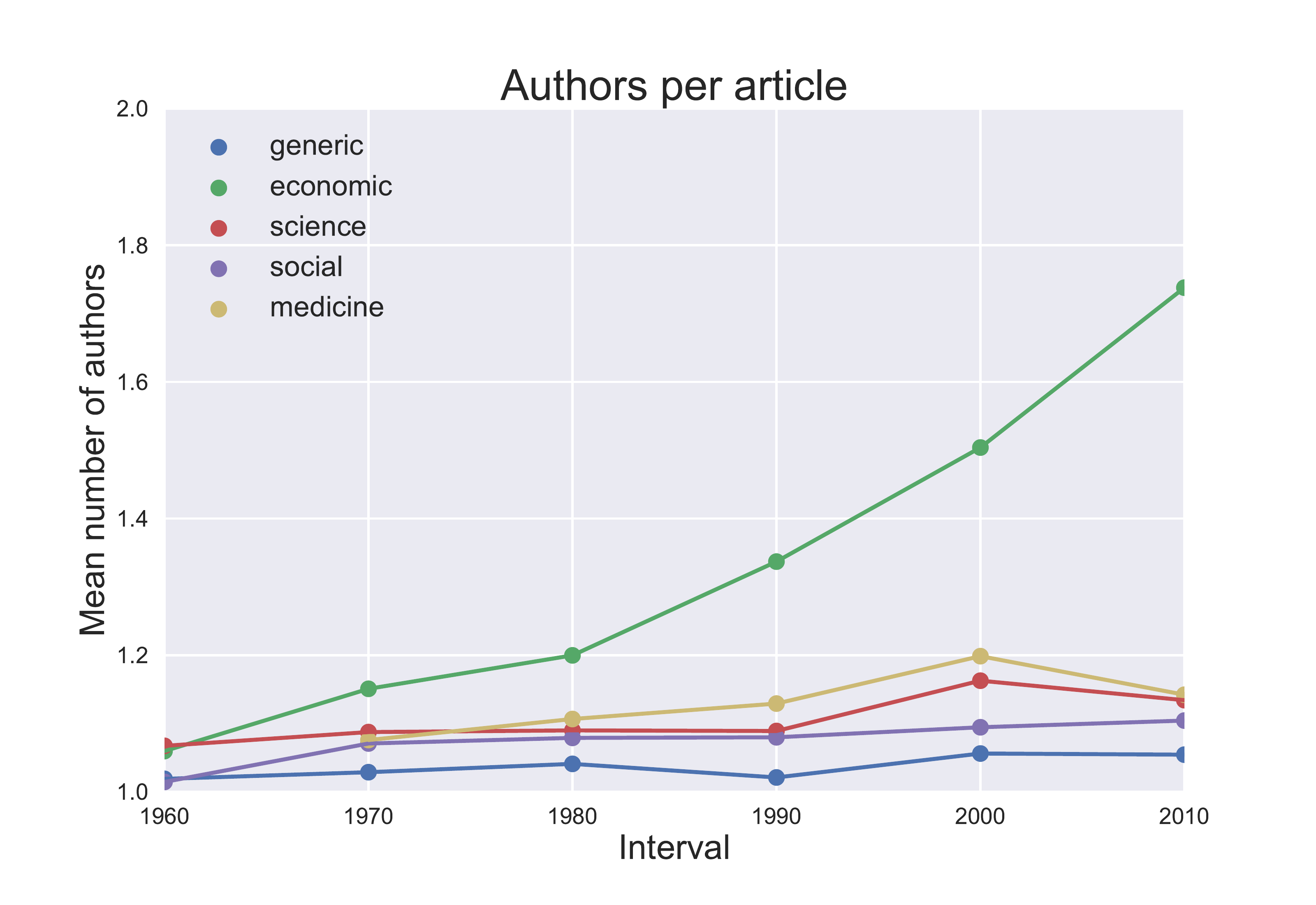}
		\subcaption{Number of authors per article.}\label{fig:7_n_auth}
	\end{minipage}
	\caption{The number of articles and number of authors per article, over specialisms and periods. Every period is individuated by its start date (the first period in fact starts in 1956). The datapoints are plotted as circles, while the lines uniting them are just meant to aid the reader.}\label{fig:7_auth_art}
\end{figure}

An important element to consider in order to compare the connectivity of networks over time is their relative similar size. The number of articles per specialism is given in Figure \ref{fig:7_n_art}, showing how from the second period, the number of articles is relatively stable within each specialism (with the minor exception of the second period for the history of medicine). A similar comparability has to hold for the number of authors and especially for the relative importance of co-authorships. Co-authorship often determines edges of weight one in the author bibliographic coupling network, therefore it has a strong influence on connectivity. As shown in Figure \ref{fig:7_n_auth}, co-authorships are very rare and stable for all specialisms, with the important exception of economic history, where they significantly raise over time. Interestingly, economic history is strongly related to economics and the social sciences, and might share some traits with these communities \citep{henriksen_rise_2016}. Connectivity results for author networks will therefore be given with and without economic history, to account for its different behavior in this respect.

\begin{figure}
	\begin{minipage}{0.5\textwidth}
		\centering\includegraphics[width=\textwidth]{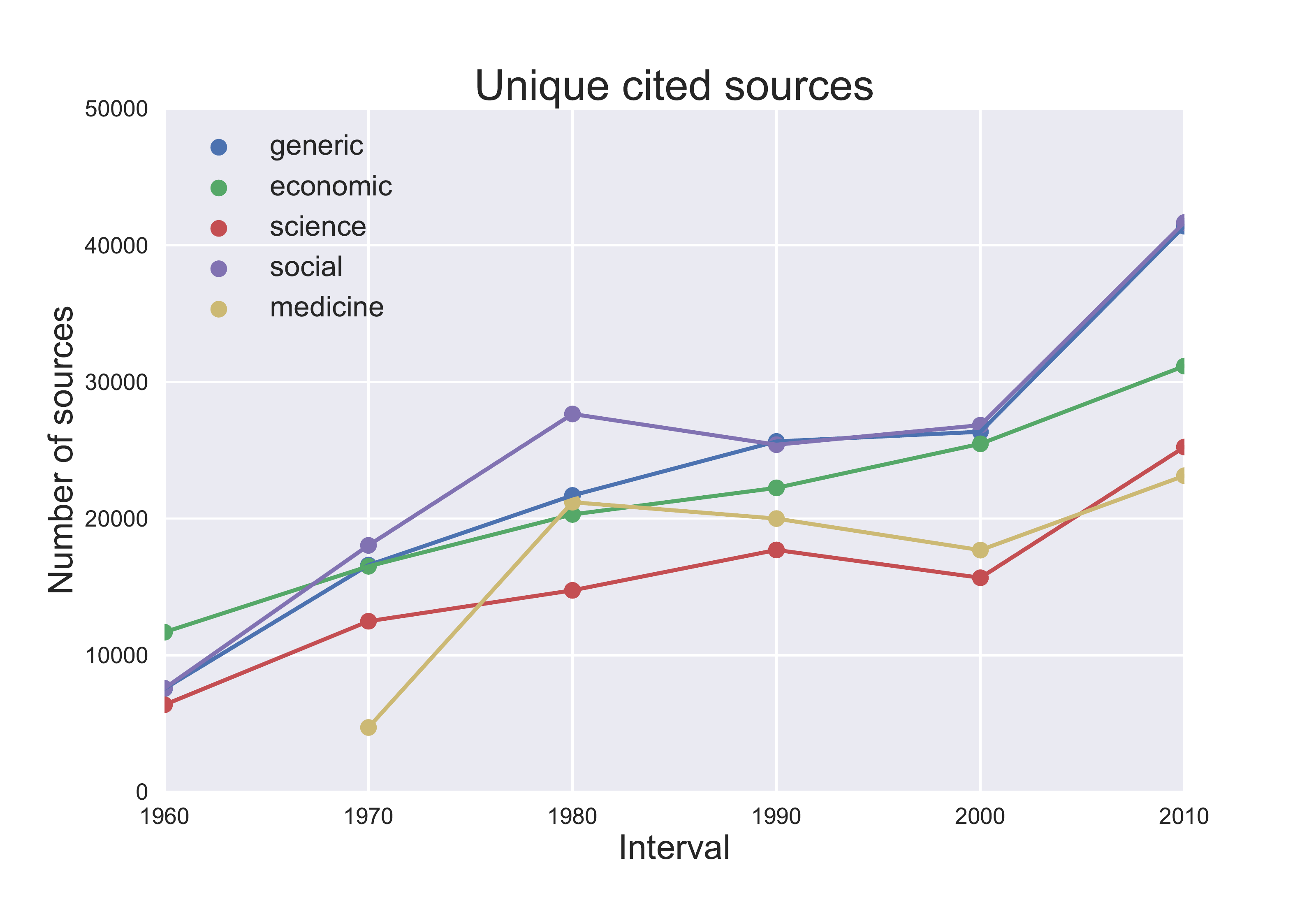}
		\subcaption{Unique cited sources per specialism.}\label{fig:overload_fields}
	\end{minipage}\hfill
	\begin{minipage}{0.5\textwidth}
		\centering\includegraphics[width=\textwidth]{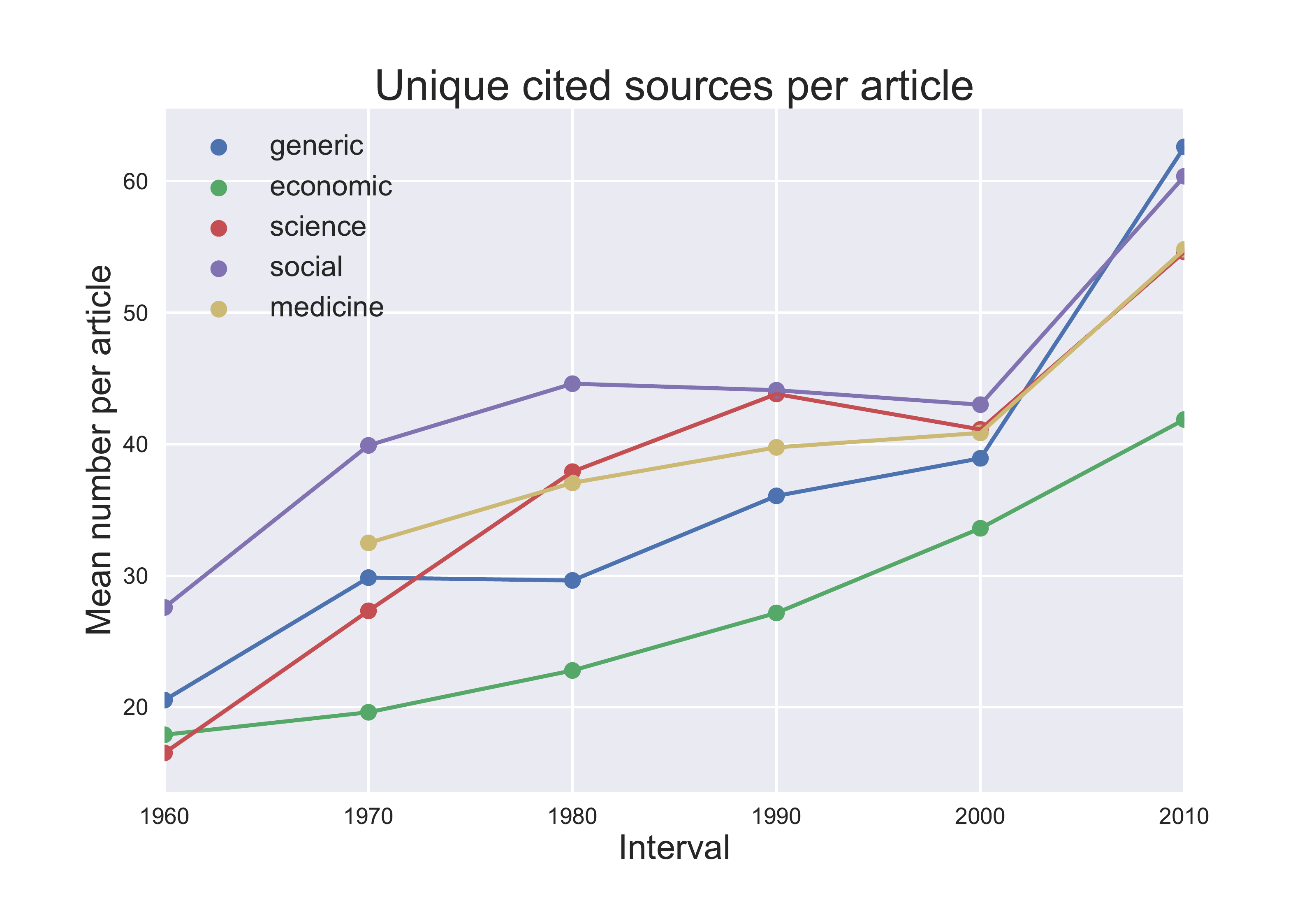}
		\subcaption{Unique cited sources per article.}\label{fig:overload_articles}
	\end{minipage}
	\caption{The number of unique cited sources per specialism and per article. Every period is individuated by its start date (the first period in fact starts in 1956). The datapoints are plotted as circles, while the lines uniting them are just meant to aid the reader.}\label{fig:7_overload}
\end{figure}

A second aspect to consider is the global number of unique references and especially the mean number of unique references per article, both given in Figure \ref{fig:7_overload}. They show a clear trend towards a progressive increase, especially pronounced over the last period (2010-2016), shared with some differences among all specialisms. If the length of reference lists raises over time, this can have an influence over the distribution of the cosine similarities among articles, as per Eq. \ref{eq_cosine}. In particular, if the raw number of shared references remains constant but the length of reference lists raises, the cosine similarity will become smaller. Therefore there are two ways for the cosine similarity to lower: when there are fewer shared references over comparably long reference lists, and when the number of shared references remains stable or anyway raises less rapidly than the length of reference lists.

\begin{figure}
	\centering\includegraphics[width=10cm]{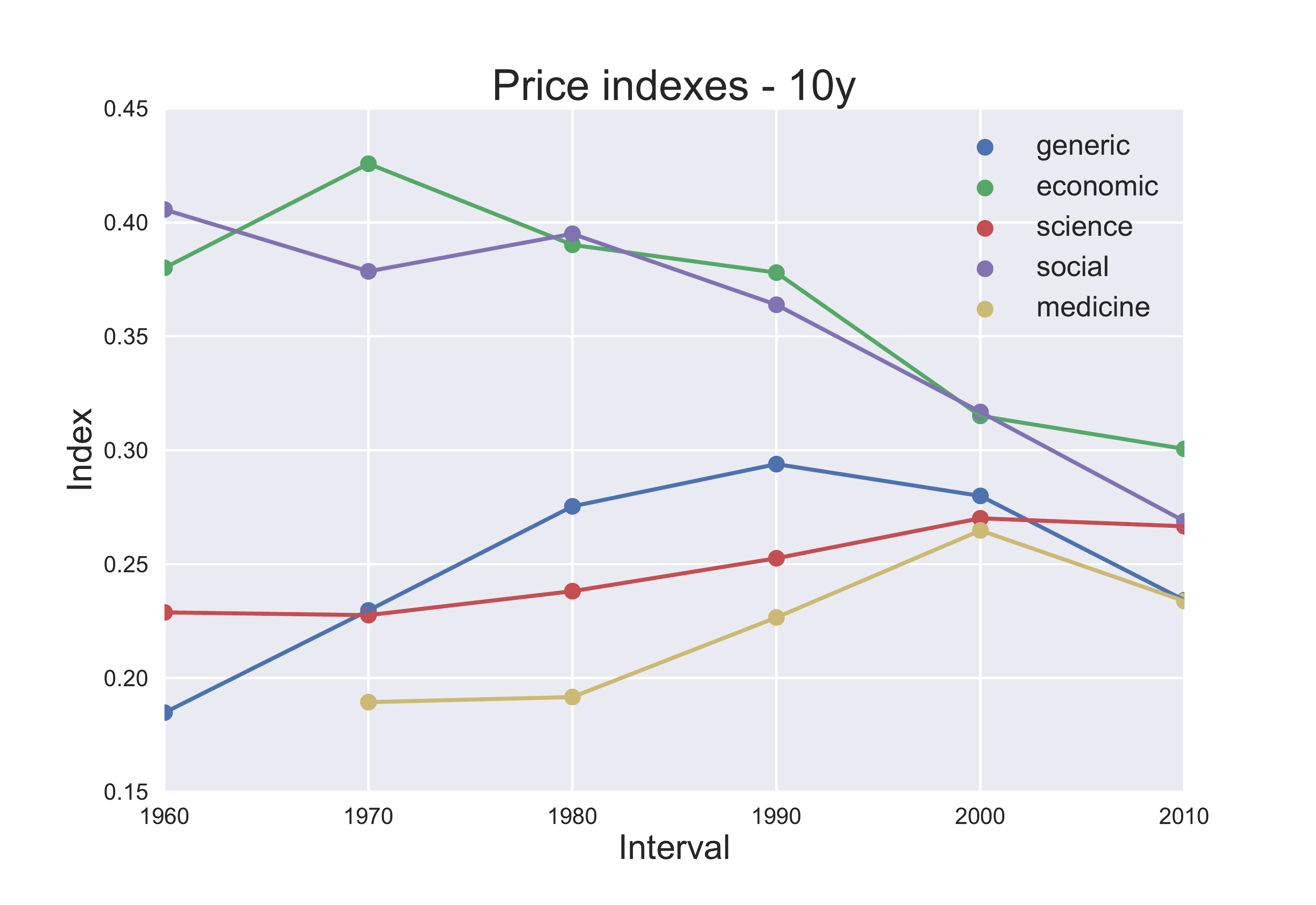}
	\caption{The Price index, over specialisms and periods. Every period is individuated by its start date (the first period in fact starts in 1956). The datapoints are plotted as circles, while the lines uniting them are just meant to aid the reader.}\label{fig:7_price}
\end{figure}

A last element worth considering is the Price index \citep{de_solla_price_citation_1970}, given in Figure \ref{fig:7_price}. The Price index is here the proportion of cited sources published maximum within 10 years from the cited one, and conveys an idea of the age of the cited literature of a specialism. Two specialisms, social and economic history, starting from a relatively high Price index in the 1950s to 70s, have been rapidly falling as their literature grew old and numerous. Conversely, the other three specialisms have seen a raise of relative stability of their Price index, which was sensibly lower to begin with.

\section{Results}\label{sec:results}

We are interested to see whether the general connectivity of the bibliographic coupling networks of different specialisms in history is changing over time. Let us thus consider Eq. \ref{eq1}, averaged over specialisms and plotted at increasing thresholds considering contiguous time periods. As a reminder, by increasingly removing edges whose weight is below the given threshold, the network collapses into many small connected components. We consider the speed of collapse in order to see whether the connectivity of these networks is declining over time.

\begin{figure}
	\centering\includegraphics[width=10cm]{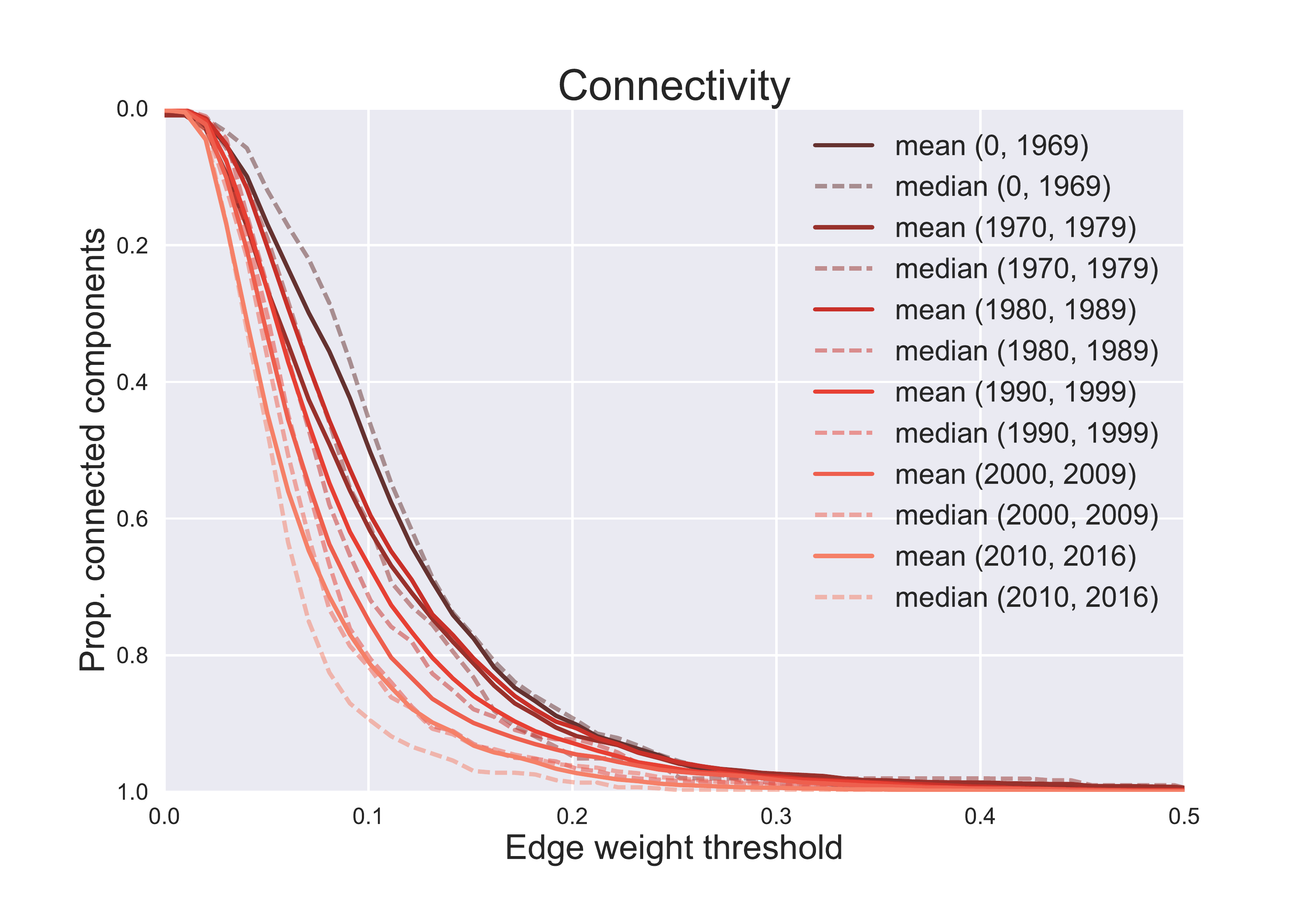}
	\caption{The connectivity of reference overlap bibliographic coupling networks over time. Please note the y axis goes from 1 to 0.}\label{fig:7_ref_conn}
\end{figure}

Starting with the article reference overlap bibliographic network (Eq. \ref{eq_cosine}), results are shown in Figure \ref{fig:7_ref_conn}. It can be clearly seen that progressively, from the 1980s at least, the mean and median connectivity is declining. Intuitively, from the first period to the last, the average bibliographic coupling network considered at threshold $0.1$ collapses from 40\% to 80\% of the total possible connected components. These results are also remarkable for showing a gradual decline in connectivity, as if this process were not driven by conjunctures or specific events, but by a steady change. This suggests that, at the article level, the decline in connectivity might be due both to a decline in the raw number of shared references and a raise in the number of non-shared ones. Results at the individual specialism level, which are here omitted for brevity, highlight that general, social and science history show a more marked decline in connectivity, whilst economic history and the history of medicine are somewhat more resilient to it, albeit participating in the same general trend.

\begin{figure}
	\begin{minipage}{0.5\textwidth}
		\centering\includegraphics[width=\textwidth]{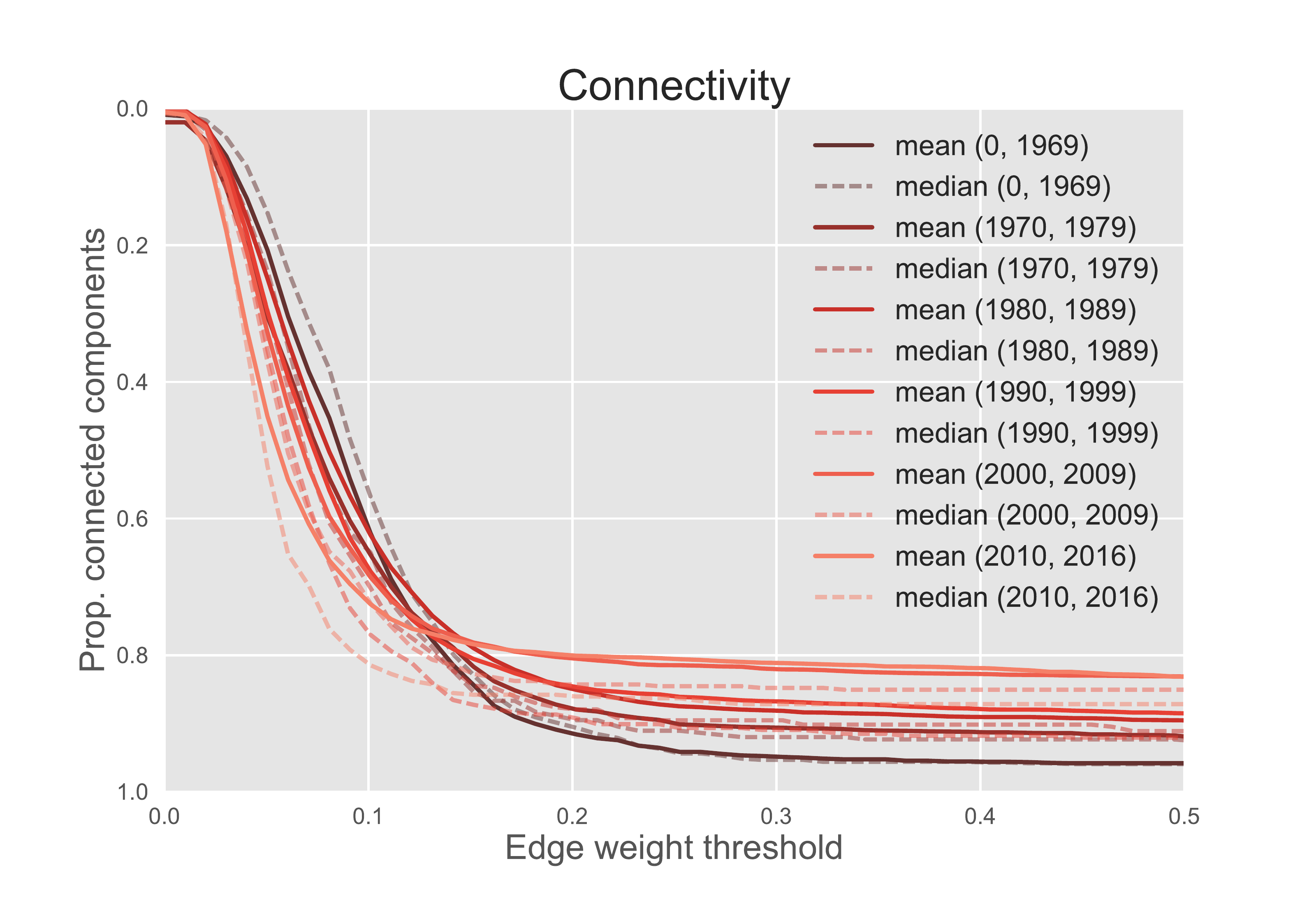}
		\subcaption{With all specialisms.}\label{fig:7_auth_conn_we}
	\end{minipage}\hfill
	\begin{minipage}{0.5\textwidth}
		\centering\includegraphics[width=\textwidth]{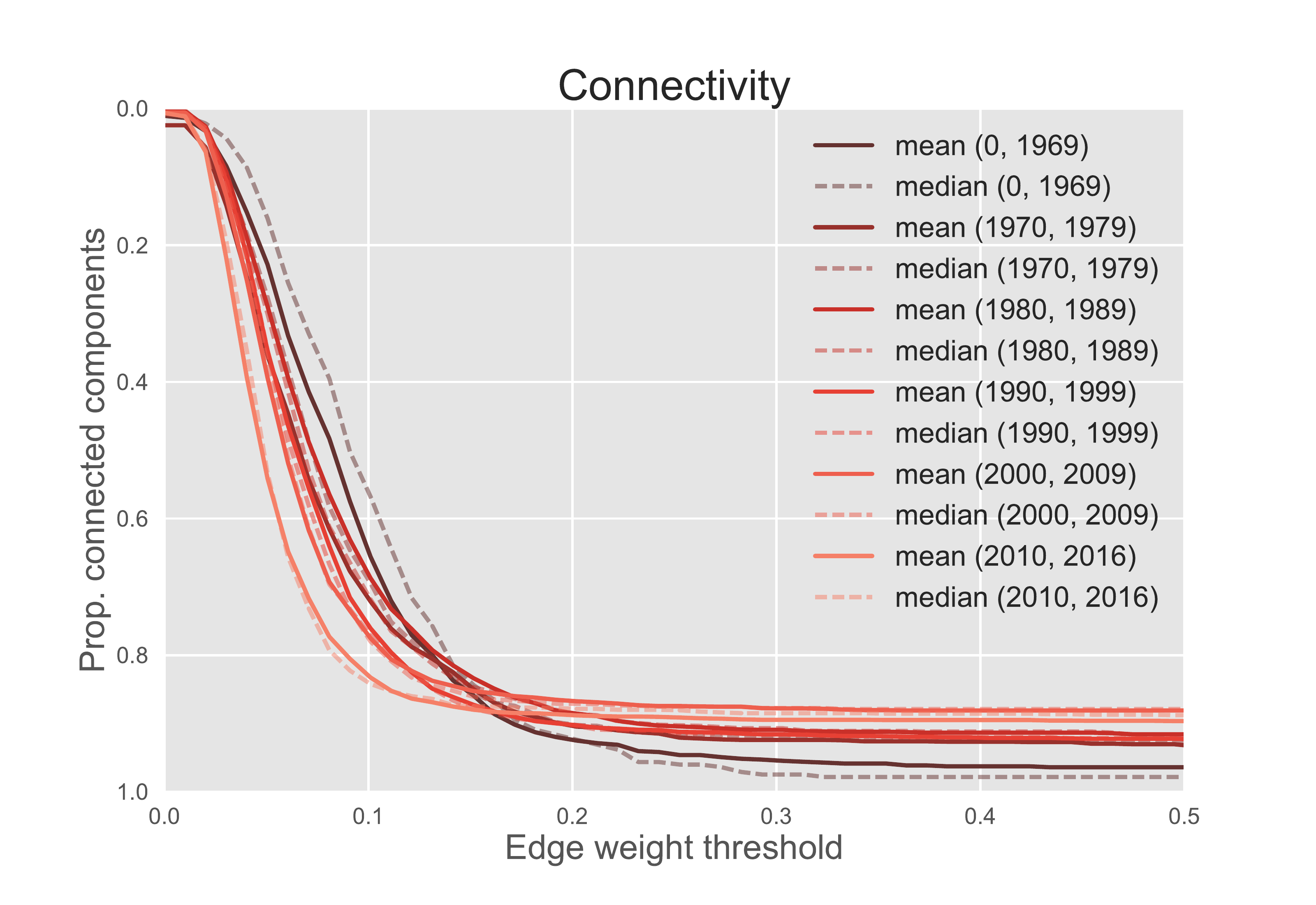}
		\subcaption{Without economic history.}\label{fig:7_auth_conn_woe}
	\end{minipage}
	\caption{The connectivity of author reference overlap bibliographic coupling networks over time. Please note the y axis goes from 1 to 0.}\label{fig:7_auth_conn}
\end{figure}

We then consider author bibliographic coupling networks, where authors are the nodes and connections are established if they share references among themselves. We account for results with and without economic history separately, due to the raising frequency of co-authorships in this domain, as previously discussed. Results are given in Figure \ref{fig:7_auth_conn}. We can see that results follow alongside article reference overlap networks, albeit with less marked effects. The decline in connectivity is particularly strong during the last time period under consideration (2010-2016), where the number of unique cited sources has been raising significantly (cf. Figure \ref{fig:7_overload}). At the author level therefore, we might be witnessing the effects of a stability in the raw number of shared references, and a rise in the number of references made. Essentially, authors seem to refer to more, non-shared sources. Finally, we can appreciate how important the impact of co-authorships is on the general connectivity of an area of research. Economic history stands out in this respect, but the general effect is still present when discarding its signal, as per Figure \ref{fig:7_auth_conn_woe}. Results at the specialism level show how economic history and the history of medicine have stable or even slightly raising connectivity at the author level, contrary to the rest of the specialisms under consideration.

\begin{figure}
	\centering\includegraphics[width=10cm]{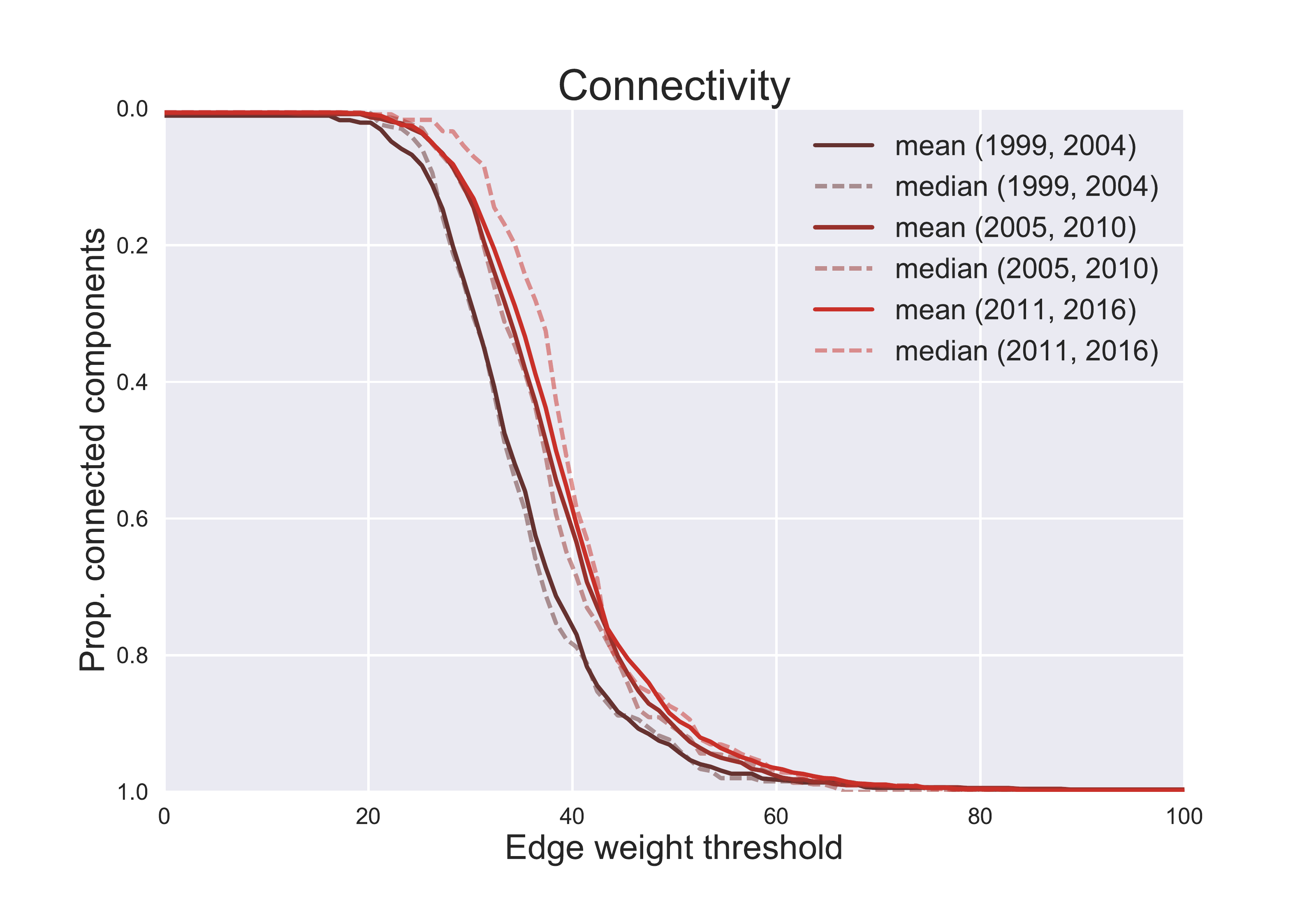}
	\caption{The connectivity of textual similarity bibliographic coupling networks over time. Please note the y axis goes from 1 to 0.}\label{fig:7_text_conn}
\end{figure}

Lastly, textual similarity bibliographic coupling networks are considered in Figure \ref{fig:7_text_conn}. We remind that these results only consider the title and abstract as text representation for an article, using data for only one journal per specialism (cf. Table \ref{tab:dataset}) and considering only the most recent time periods. Interestingly, results point to a stable or raising connectivity at this level. All journals participate in this trend, with Isis showing a particularly strong raise in similarity from the first to the second period.

In summary, we have shown that the general connectivity of article reference overlap bibliographic coupling networks is gradually and steadily declining over time. The author reference overlap connectivity is also in general declining, but only slightly, with some exceptions and an important counter effect provided by the raise in the number of co-authorships, particularly marked for economic history. Lastly, the textual similarity connectivity appears to be stable over recent times.

\section{Conclusions}\label{sec:conclusions}

In this article we explored the extent to which the intellectual landscapes of different specialisms in history are changing over time. The proposed method was that of exploring the connectivity properties of bibliographic networks of articles and authors constructed using reference overlap and textual similarity. The limitations of this study include, and are probably not limited to, an the exclusive focus on history and on journal articles as citing publication typology. Indeed, our conclusions hold to the extent of these significant limitations. Nevertheless, our results highlight some aspects of importance. 

The connectivity of article bibliographic coupling networks is declining over time, and this can be due either to a decline in the raw number of shared sources or to the raising number of non-shared ones, possibly both. The connectivity of author bibliographic coupling networks is also declining but less rapidly, in part due to the slight counter effect of co-authorships, which are increasing especially in economic history. It is also evident that author networks decline in connectivity more markedly over the last period, when a sensible raise in the number of unique references is recorded (i.e. reference lists become longer). In accordance to what has been found for the sciences and social sciences, the concentration of citations might be lowering in recent times in part due to the raising amount of references made per article. This phenomenon might explain the reported trend, albeit previous evidence points to a general stability in the concentration of citations for the humanities \citep{lariviere_decline_2009}. Lastly, the persisting levels of connectivity in the textual similarity networks might highlight a steady effort in integrating research results relying on shared vocabularies or even narratives.

Previous studies have shown how historians are not narrowing the scope of their attention with respect to primary sources, but are indeed broadening their topical and methodological perspectives \citep{colavizza_hoh_2018}. The results of this chapter can therefore be tied together into a possible interpretation: the undeniable emergence of various directions of research in historiography over recent decades, coupled with a growing amount of published results, is determining a gradual decline of the citation connectivity of publication networks as scholars produce more fragmented or specialized research. Nevertheless, at the level of scholars this decline is only marginal, thanks in part to a generally modest but in some specialisms more significant raise in collaborations as evidenced by co-authored results. Another, possibly more important way for historians to tie their results together is their reliance on shared vocabularies. More work is necessary to further elucidate this hypothesis, and more generally to complement citation with text data in bibliometrics.

History thus emerges as a discipline where the pressures of growth are determining i) a decreasing reference connectivity of publications within specialisms and, to a lesser degree, of scholars, ii) a perhaps only too tenuously growing propensity for collaborations and, we speculate, iii) a constant reliance on shared narratives in order to integrate new results together. These results highlight an emergent property of the intellectual organization of history, faced with the gradual growth of accumulated knowledge. According to \cite{jones_burden_2009}, the accumulation of knowledge leads to an increasing educational burden, a narrowing of expertise (raise in specialization) and an increased propensity for teamwork within a cohesive social system. Importantly, these general strategies can differ in magnitude cross-sectionally, likely in a way which is correlated to the degree of consensus within the specialism or even the field, as we briefly discussed in the introduction. Considering the scholarly output of historians, we can distinguish between interpretive work and work on primary evidence (such as catalogs, critical editions or databases) \citep{ziman_public_1968}. Whilst the latter might lend itself more easily to consensus on methods and results, thus possibly going further into exploring specialization and teamwork within a shared methodological framework, the former is likely to put a premium on novelty and originality, defined in a variety of ways \citep{guetzkow_what_2004}. This attitude to the reception of interpretive work is also evidenced by the tendency of humanities authors to take explicit points of view and emphasizing their own contributions \citep{hyland_disciplinary_2006}. As a result, `citation fragmentation' of interpretive works (such as most journal articles) might indeed occur, but not been caused by specialization nor paralleled by a growing propensity for teamwork. Yet if intellectual fragmentation might slow scientific progress by limiting scholars' influence among their peers \citep{balietti_disciplinary_2015}, in the humanities the integration of new results into the body of disciplinary knowledge might simply be more reliant on the principal form of expression in these fields: narratives.

\bibliographystyle{abbrvnat}
\bibliography{bib.bib}   

\end{document}